\documentclass[11pt]{article}

\usepackage{amsmath,graphicx}
\usepackage{multirow}
\usepackage{amsfonts}
\usepackage{amssymb}
\usepackage{amscd}

\def\hybrid{\topmargin -20pt    \oddsidemargin 0pt
        \headheight 0pt \headsep 0pt
        \textwidth 6.25in       
        \textheight 9.5in       
        \marginparwidth .875in
        \parskip 5pt plus 1pt   \jot = 1.5ex}

\hybrid

\numberwithin{equation}{section}
\numberwithin{table}{section}
\renewenvironment{thebibliography}[1]
 { \small
   \begin{list}{[\arabic{enumi}]}
    {\usecounter{enumi} \setlength{\parsep}{0pt}
     \setlength{\itemsep}{5pt} \settowidth{\labelwidth}{#1.}
     \sloppy
    }}{\end{list}}
 
\newcommand{\be}{\begin{equation}}
\newcommand{\ee}{\end{equation}}
\newcommand{\bea}{\begin{eqnarray}}
\newcommand{\eea}{\end{eqnarray}}
\newcommand{\ba}{\begin{array}}
\newcommand{\ea}{\end{array}}
\newcommand{\bc}{\begin{center}}
\newcommand{\ec}{\end{center}}

\def\a{\alpha}
\def\b{\beta}
\def\g{\gamma}
\def\l{\lambda}
\def\d{\delta}
\def\e{\epsilon}
\def\t{\theta}

\def\L{\Lambda}

\def\p{\partial}

\newcommand{\sym}[2]
   {\omega\!\left(#1,#2\right)}
\newcommand{\symr}[2]
   {\omega_\rho\!\left(#1,#2\right)}
\newcommand{\symJ}[2]
   {\omega_J\!\left(#1,#2\right)}

\newcommand{\cref}{{\bf [check ref]}}

\begin{document}


%
%


\rightline{\small CERN-PH-TH/2005-199}
\rightline{\small DISTA-UPO-05}
\begin{center}

\vskip 2cm
{\Large \bf N=2 Superparticles, RR Fields, }\\
\vskip 0.4cm 

{\Large \bf and Noncommutative Structures of (super)-Spacetime}\footnote{Based on a lecture given at 43$^{rd}$ International School of Subnuclear Physics, Erice, Sicily, Italy, Aug. 2005.}

\vskip 1.5cm

{\bf Pietro Antonio Grassi}

\vskip 0.8cm
{\em Centro Studi e Ricerche E. Fermi, 
Compendio Viminale, I-00184, Roma, Italy,}\\
\vskip 0.2cm
{\em CERN, Theory Unit, Physics Department, 
1211-CH Geneva 23, Switzerland, 
}\\
\vskip 0.2cm
{\em DISTA,
Universit\`a del Piemonte Orientale,
via Bellini 25/g, I-15100, Alessandria, Italy. 
}\\
\vskip 0.4cm
{\tt pietro.grassi@cern.ch}
\vskip2cm

\today

\begin{abstract}

\vskip0.6cm

The recent developments in superstring theory prompted 
the study of non-commutative structures in superspace. 
Considering bosonic and fermionic strings 
in a constant antisymmetric tensor background 
yields a non-vanishing commutator between the bosonic 
coordinates of the spacetime. Likewise, the presence of constant 
Ramond-Ramond (RR) background leads to a non-vanishing 
anti-commutator for the Grassmann coordinates of the superspace. 
The non-vanishing commutation relation between bosonic coordinates can also 
be derived using a particle moving in a magnetic background, we use 
N=2 pure spinor superparticles and $D0$-branes to show how the non-commutative structures emerge in superspace. It is argued how a
$D0$-brane in a background of RR fields reproduces the results 
obtained in string theory.   

\end{abstract}
\end{center}
\pagebreak
{\small\tableofcontents }
\vskip .5cm
\hrule

\section{Introduction}
\label{intro}
During the last years, several new ideas emerged from 
the marriage of non-commutative geometry to quantum field 
theory and string theory. This is due to the discovery that the spacetime 
generated by strings propagating on a non-trivial background 
\cite{Chu:1998qz,Chu:1999gi,Ardalan:1998ce,Seiberg:1999vs,Bigatti:1999iz}
is non-commutative. Furthermore string theory 
provides a meaningful way to construct 
quantum field theories on non-commutative spaces. 

Nevertheless 
the history of non-commutative geometry and non-commutative 
structures of spacetime is definitely longer and it has its roots in quantum 
mechanics. Indeed, it was first realized by Peierls using 
non-relativistic quantum mechanics that the motion of a charged particle in 
presence of non-trivial external magnetic fields can be described 
by a free Hamiltonian assuming non-vanishing commutation relations 
for the coordinates \cite{Peierls:1933aa,Dunne:1989hv,Jackiw:2001dj}. 

The simplest example of non-commutative 
spacetime \cite{Snyder:1946qz,Yang:1947ud} 
is represented by the Heisenberg algebra 
of the coordinates 
\be\label{IA}
[x^{m}, x^{n}] = i \, \theta^{mn}
\ee
where $\theta^{mn}$ is a constant antisymmetric tensor. Based on this 
prototype, it has been developed an enormous amount of new mathematics 
which we are not going to review here. 
The commutation relations (\ref{IA}) can be recovered from 
the Hamiltonian of a charged particle moving in a background of magnetic field 
in the massless limit $m \rightarrow 0$.  In that limit the particle 
is confined in the lowest Landau level and this limit can consistently be taken 
if some constraints on the momenta are imposed. These are second-class 
constraints that have to be treated using the Dirac brackets and this 
yields the commutation relations (\ref{IA}). We briefly review 
this model in Sec.~\ref{particle}. Then, we move to the supersymmetric version. 

As is well-known, bosonic particles and bosonic string theories are not 
sufficient to provide a complete description of particle physics since 
they do not include fermionic degrees of freedom (except maybe only 
for some unphysical ghost fields). Thus, we have to extend 
the bosonic theory to a fermionic one. There are essentially two ways to do it: 
adding some fermionic (anticommuting) {\it worldline} spinors $\psi^{m}$ 
(or worldsheet spinor 
in the case of superstrings) or adding some fermionic {\it target-space} spinors $\theta^{\a}$
\cite{Casalbuoni:1976tz,Brink:1980nz}. In the former 
case supersymmetry on the worldsheet has to be imposed for a consistent formulation of the model, 
whereas for the latter case, one has to impose a new gauge symmetry, known as $\kappa$-symmetry 
\cite{Siegel:1983hh}, and this leads to supersymmetry in the target space. 
We recall the basic ingredients of this superparticle model in Sec.~\ref{superparticles}.

The quantization of superparticle is unfortunately very problematic. 
The action is obtained from the bosonic 
one by replacing the momentum $\Pi^{m}$ with its supersymmetric version 
$\Pi^{m} = \dot x^{m} + \theta \gamma^{m} \dot \theta$ and this leads to fermionic constraints since the 
momentum $p_{\a}$ is algebraically related to its conjugated variable $\theta^{\a}$.  
However, these constraints mix first-class constraints -- which generate the $\kappa$-symmetry -- with 
second-class constraints and there is no Lorentz-covariant way to separate the twos. 
Several procedures were conceived  to covariantly quantize these models (see for example 
\cite{Grassi:2000qs} and the references therein), but most of them were nonpractical for 
computations and were abandoned.\footnote{It is important to mention that the superparticle 
can be quantized using the light-cone gauge. In that case the spectrum can be easily computed and 
tree level computations can be performed \cite{Green:1987mn}. However, there are several limitations to go beyond this point because of lacking of Lorentz covariance.} 
 
On the other hand, the recent work 
by N. Berkovits \cite{Berkovits:2000a} provides a new technique 
to handle the quantization of the superparticle and the superstring theory. 
In this new framework, the action of the superparticle is replaced by 
a free action and the physical states are constructed using a BRST charge 
acting on the Hilbert space of free fields. To be more precise, 
some of fields are not really free. Indeed, to define a BRST charge $Q$, 
one has to introduce new degrees of freedom which play the role of ghosts here 
denoted by $\l^{\a}$, and the nilpotency of $Q$  implies 
the quadratic constraints
\be\label{ps1}
\l^{\a} \g^{m}_{\a\b} \l^{\b} = 0\,, 
\ee
where $\gamma^{m}_{\a\b}$ are the Dirac matrices in the Majorana basis and 
they are symmetric (in 10 dimensions). The spinors satisfying eq.~(\ref{ps1}) 
are known as {\it pure spinors} and the formalism is now denoted as 
Pure-Spinor Formulation. 
Here, we consider only the Pure Spinor formulation of superparticle and $D0$-branes and 
we refer to \cite{Berkovits:2001rb,Berkovits:2002d,Berkovits:2002uc}. 
It has been also considered 
the possibility to remove the constraint by adding new ghost fields in the 
work \cite{Grassi:2001,Chesterman:2002,Aisaka:2002sd} and 
in the specific case of superparticle this was explored in \cite{Grassi:2002xf}. However
for the purposes of the present work we will use the Pure-Spinor formulation 
whose basic 
ingredients will be reviewed in Sec.~\ref{quantization}. Thus, given a consistent way to 
quantize the superparticle we can study the spectrum and the interactions. 

At the massless (lowest) level string theory can be described 
by an effective theory of supergravity and the spectrum 
consists of  a bosonic sector with the graviton 
$G_{mn}$, the NS-NS antisymmetric tensor $B_{mn}$ , the dilaton 
$\phi$ and a set of $p$-forms $F_{p}$\footnote{In the case of N=2 d=10 supergravity 
there are two possibilities: type IIA with $F_{2}, F_{4}$ and type IIB 
with $F_{1}, F_{3}, F^{+}_{5}$ (where the last form is selfdual).}, and 
a fermionic sector, a.k.a. NS-R or R-NS sector, which contains the gravitinos 
$\Psi^{\a}_{m}$ (see \cite{Polchinski:1998rq} for a complete reference). 
The interest of superparticles in this context is due to the fact that 
they can be viewed as truncations of string theory to the massless sector. Therefore these models 
are useful to deduce some general aspects of string theory such as the spectrum of the massless modes, 
their equations of motion, and some radiative corrections, 
even if they can be used only a limited amount of amplitude computations \cite{Green:1997as,Anguelova:2004pg}.

To be more precise, the N=1 d=10 superparticle describe the multiplet of  N=1 super-Yang-Mills theory. 
The spectrum is characterized by the gluon (8 on-shell dofs) and the gluino (8 fermionic dofs). 
It is formulated in the superspace, but there are no auxiliary fields since the multiplet is on-shell. 
An N=2 superparticle in 10 dimensions describes the on-shell modes of N=2 supergravity, namely 
64 bosonic and 64 fermionic degrees of freedom.

Whereas string theory can be consistently formulated only in 10 dimensions, superparticle 
models can be formulated also in lower dimensions.\footnote{Recently, it has been 
discovered that also superstrings can be formulated in lower dimensions 
\cite{Berkovits:2005c,Grassi:2005sb,Wyllard:2005fh,Chandia:2005fi} and these models 
can be viewed either as the uncompactified part of 10 dimensional superstrings or as the non-Liouville sector of non-critical superstrings \cite{Grassi:2005kc}.} These models 
in lower dimensions are easier to be used since the BRST conditions for physical states 
does not put the theory on shell and there is a wider range of consistent backgrounds (vacua). 

For what concerns the interactions we have to recall that 
the superparticle as well as superstrings couple to their own 
background. This means, for instance, that N=2 d=4 superparticle couples to 
N=2 d=4 supergravity. In Sec.~\ref{background} this is described in detail. 
More important, we have to underline that N=2 d=4 supergravity is characterized 
by a graviton, two gravitinos and a RR field (known as graviphoton in the 
literature) and the coupling with N=2 d=4 superparticle is dictated by the BRST symmetry 
In addition, since for the d=4 model the supergravity does not need to be 
on-shell we can choose to set to zero all background fields except the RR field. 

We will show later how the RR fields 
lead to deformations of anticommutative structure of 
superspace. But before describing this result, it is worth to say few words 
about superspace. 

Let us remind the reader that {\it superspace} \cite{Gates:1983nr} 
is a powerful technique to handle supersymetric 
theories, it is characterized by the bosonic coordinates of manifold and a set of Grassmann coordinates 
in the spinor rapresentation of the Lorentz group. The superspace technique provides {\it 1)} a 
very compact way to write the equations of motion for the entire supersymmetric multiplet, {\it 2)} 
an extremely economic way to compute Feynamn diagrams taking into account supersymmetry and, 
{\it 3)} a guideline to construct effective actions of supersymmetric theories. Finally, superspace 
is naturally embedded in the Pure Spinor formulation of string theory.  

Now, we are finally in the position to study the supersymmetric analog of noncommutative 
geometry of bosonic theory (\ref{IA}). At this time, we study the deformation of the anticommutator 
between fermionic coordinates $\theta^{\a}$ \cite{Ferrara:2000mm,Klemm:2001yu}. 
We have to recall that there are 
several studies in that direction \cite{Brink:1981nb,Schwarz:1982pf,Bars:2003dq} where 
the second-class constrained were used to show that there is a fundamental 
non-commutative superspace in the quantization of superparticle. This is reviewed in 
this new formulation, for two reasons: {\it 1)} the pure spinor quantization method is the 
only consistent way to quantize superparticle without losing the super-Poincar\'e invariace; {\it 2)} 
it is show the role of the RR fields in the present analysis in Sections~\ref{RRI} and \ref{RCor}.  

However, only recently \cite{deBoer:2003dn,Ooguri:2003qp,Berkovits:2003kj}, 
using the pure spinor formulation of string theory 
it is shown how the RR fields deform the anticommutation relations as follows
\be\label{ac}
\{ \theta^{\a} , \theta^{\b} \} = \alpha'^{2} F^{\a\b}\,, 
\ee
where $F^{\a\b}$ are the RR fields of N=2 supergravity. It is discussed the 
implications of these new anticommutator relations on quantum field 
theory in paper \cite{Seiberg:2003yz} and in the several papers that followed it. 
The relation (\ref{ac}) can also be derived in the context of quantum mechanics, or better 
in the context of superparticle and this is the purpose of this note. 
We show that in order to reproduce the relation (\ref{ac}) we have to use a peculiar 
type of superparticle known as $D0$-brane. It is used again the pure spinor formulation 
and it is shown that the quantization of the D0 brane leads 
to non-commutativity in the 
superspace. 

Here, we also discuss the perturbation theory and the limit where 
the RR fields can be taken either very weak or very strong. Some interesting 
results emerge from this preliminary analysis and further developments will be discussed elsewhere \cite{Grassi:2005gt}.  

Finally, we would like to point out that after the original papers \cite{deBoer:2003dn,Ooguri:2003qp}, numerous studies followed them and there is now 
a wide literature on the applications. Since we are not discussing 
the applications to gauge theories or the geometrical implications of the 
deformed commutation relations in bosonic and fermionic case, we will 
not include this part of the literature in our references. 


\section{A Particle in a Magnetic Field}
\label{particle}

We briefly review the formulation of a non-relativistic massive and charged 
particle in presence of 
magnetic field. We will do it in a generic dimension and the position of the 
particle is described by its coordinates 
$x^{m}$ (with $m=1, \dots, d$). We introduce a constant 
background $B_{mn} = - B_{nm}$, thus, the action reads ($\dot A = \p_{\tau}A$) 
\be\label{ppA}
S = \int d\tau \Big( \Pi_{m} \dot x^{m}  - {1\over 2 m} \Pi_{m} \Pi^{m} + 
B_{mn} \dot x^{m} x^{n}
\Big)\,.
\ee
The indices are raised and lowered with the flat metric $\eta_{mn}$. 
The conjugate momentum $P_{m}$ can be easily computed and 
it gives $P_{m} = \Pi_{m} + B_{mn} x^{n}$. Now, 
if we impose the quantization rules
$[P_{m}, x^{n}] = i \delta_{m}^{~n}$, we have that 
\be\label{ppB}
[x^{m}, x^{n}]= 0\,, ~~~
[\Pi_{m}, x^{n}] = i \delta_{m}^{~n}\,, ~~~
[\Pi_{m}, \Pi_{n}] = i B_{mn}\,.
\ee
The other equations of motion are 
\begin{eqnarray}\label{ppC}
m \,  \dot x^{m} = \Pi^{m} \,, ~~~~ \dot \Pi^{m} + B^{m}_{~n} \dot x^{n} =0\,.
\end{eqnarray}
For this equations, it follows that $\dot P_{m}=0$. Therefore, 
in order to take the limit $m \rightarrow 0$, we have to impose the constraints 
\be\label{ppCA}
\Pi^{m} \approx 0\,,
\ee
which are second-class constraints. They have to be treated using the 
Dirac brackets (see \cite{Henneaux:1992ig} for the definition of Dirac brackets) and 
this leads to the non-commutation relations for the coordinates 
\be\label{ppD}
[x^{m}, x^{n}]_{D} = (B^{-1})^{mn}\,,
\ee
where the subscript denotes the Dirac brackets. 

This model is interesting for three aspects: {\it 1)} it represents 
a simple solvable model of a particle moving in a non-trivial background; {\it 2)} 
it gives the non-commutative relations between the coordinates of the spacetime and, 
finally {\it 3)} it requires Dirac brackets for its quantization. All these ingredients 
will be found again in the subsequent sections.  


\section{Superparticles and the BRST symmetry}
\label{superparticles}

\subsection{Action and $\kappa$-symmetry}
\label{action}

We use Dirac basis for gamma matices, and the spacetime is taken to be 4 dimensional. 
 The field content is represented by the bosonic coordinates $x^{m}$ where $m=0,\dots,3$, 
two anticommuting Dirac spinors $\t^{\a}_{L}, \t^{\a}_{R}$ 
with $\a=1, \dots, 4$ and their conjugate momenta $P_{m}, p_{L\a}$ and $p_{R\a}$. 
Since we are considering an N=2 model, we have introduced the notation $L/R$ to distinguish 
between the two flavours of the spinors. In the case of d=4, there is no distinction between type IIA/B 
since the theory is not chiral in the present case.  
The Dirac matrices $\g^{m}_{\a\b}$ are the usual $4 \times 4$ matrices
and satisfy the Fierz identities $\g^{}_{m, (\a\b} \g^{m}_{\g)\d} =0$. 

Let us consider the superparticle action 
\cite{Casalbuoni:1976tz,Brink:1980nz,Siegel:1983hh}
\be\label{CBS}
S = \int d\tau ( P_{m} \Pi^{m} - {e \over 2} P_{m} P^{m})
\ee 
in the first order formalism with
\be\label{CBSA}
\Pi^{m} = \dot{x} + {i\over 2} \t_{L} \g^{m} \dot \t_{L} + {i \over 2} \t_{R} \g^{m} \dot \t_{R}\,.
\ee 
This action is invariant under the $\kappa$-symmetry
and under the reparametrization of the worldline
$$
\d \t^{\a}_{L}  = (\not\!P \kappa_{L})^{\a}\,, ~~~
\d \t^{\a}_{R}  = (\not\!P \kappa_{R})^{\a}\,, ~~~
\d P^{m} =0
$$
\be\label{kap}
\d x^{m} = \zeta P^{m} + {i\over 2} \Big(  \t_{L}\g^{m} \not\!P 
\kappa_{L} + \t_{R} \g^{m} \not\!P\kappa_{R} \Big)\,,~~~~~~
\ee
$$
\d e = \dot\zeta + 2\, i\, \Big(\dot\t^{\a}_{L} \kappa_{L\a} + \dot\t^{\a}_{R} \kappa_{R\a}\Big)\,.
$$
where $\kappa_{L/R}$ are the infinitesimal gauge parameters of 
$\kappa$-symmetry and $\zeta$ is the parameter for diffeomorphisms. 

From the action~(\ref{CBS}), 
we deduce the fermionic constraints 
\be\label{conA}
d_{L\a} = p_{L\a} + {i\over 2} P_{m} (\g^{m} \t_{L})_{\a} \approx 0\,,
\ee
$$
d_{R\a} = 
p_{R\a} + {i\over 2} P_{m} (\g^{m} \t_{R})_{\a}\approx 0\,, ~~~~
$$
which satisfy
\be\label{conB}
\{d_{L\a}, d_{L\b}\} = P_{m} \g^{m}_{\a\b}\,,~~~~~
\{d_{R\a}, d_{R\b}\} = P_{m} \g^{m}_{\a\b}\,,~~~~~
\{d_{L\a}, d_{R\b}\} = 0\,.
\ee
The last equations are obtained using the canonical commutation relations 
$[P_{m}, x^{n}]= i \eta_{m}^{~n}$, $\{p_{L\a}, \t^{\a}_{L}\} = - i \delta_{\a}^{~\b}$, 
$\{p_{R\a}, \t^{\b}_{R}\} = - i \delta_{\a}^{~\b}$. 

We have to notice the following facts: {\it 1)} there are first- and second-class 
constraints generated by the operators $d_{L\a}$ and $d_{R\a}$, and they cannot 
be disentangled without breaking Lorentz covariance, so the technique of Dirac 
backets cannot be used here, {\it 2)} the second-class constraints of the superparticle 
have been used in \cite{Brink:1981nb,Schwarz:1982pf,Bars:2003dq} to derive non-(anti)commutation 
relations among the fermionic coordinates of the superspace. However, since as 
it stands the superparticle model cannot be quantized 
we use the Pure-Spinor Formulation. 


\subsection{Quantization}
\label{quantization}

We briefly review the Pure-Spinor formulation of superparticle 
\cite{Berkovits:2001rb,Berkovits:2002d,Berkovits:2002uc}. 

We introduce the commuting spinors 
$\l^{\a}_{L}$ and $\l^{\a}_{R}$, 
which satisfy the pure spinor conditions 
\be\label{ps}
\l_{L}\g^{m} \l_{L} = 0\,, 
~~~~~\l_{R} \g^{m} \l_{R}=0\,,
\ee 
and their conjugate momenta $w_{L\a}, w_{R\a}$.  

We define the BRST operators 
\be\label{conC}
Q_{L} = \l^{\a}_{L} d_{L\a}\,, ~~~~~~~~ Q_{R} = \l^{\a}_{R} d_{R\a}\,.
\ee
They have the usual form $ghost \times constraint$. Due to pure spinor constraints (\ref{ps}), 
they are nilpotent up to the gauge transformations of $w_{L\a}, w_{R \a}$ with 
the local parameters $\L_{L}$ and $\L_{R}$ given by 
\be\label{conD}
\delta w_{L\a} = \L_{L m} (\g^{m} \l_{L})_{\a}\,, ~~~~~~~~
\delta w_{R\a} = \L_{R m} (\g^{m} \l_{R})_{\a}\,.
\ee
These gauge transformations remove the degrees of freedom 
from the spinors $w_{L\a}$ and $w_{R\a}$ to match those of the pure spinors 
$\l^{\a}_{L}$ and $\l^{\a}_{R}$. Following the usual prescription of the BRST 
quantization rules, we can define the quantum action as follows \cite{Oda:2001zm}
\be\label{conDA}
S_{0} =  \int d\tau ( P_{m} \Pi^{m} - {1\over 2} P_{m} P^{m}) + 
Q_{L} \int d\tau w_{L\a} \dot\t^{\a}_{L} - Q_{R} \int d\tau w_{R\a} \dot\t^{\a}_{R}\,.
\ee
Even if it seems the usual BRST procedure, we have to notice that 
the BRST operators $Q_{L}$ and $Q_{R}$ are nilpotent only up to gauge 
transformations (\ref{conD}). This compensates the fact that the Brink-Schwarz superparticle 
action  (\ref{CBS}) is not invariant under the BRST transformations. In addition, we can always add to the 
action BRST invariant terms. The reparametrization is fixed by the gauge condition $e=1$, and 
we have to add the corresponding ghosts $\int d\tau b \dot{c}$. However, there is no procedure 
to get (\ref{conDA}) from an honest gauge fixing of the action~(\ref{CBS}) 
(a suggestion how this might work is given in \cite{Oda:2001zm,Aisaka:2005vn}). 

By exploiting the different contributions in (\ref{conDA}), we obtain 
\be\label{conE}
S_{0} = \int d\tau 
\Big( \dot \t^{\a}_{L}\, p_{L\a} + \dot \t^{\bar\a}_{R}\, p_{R\bar\a} + P_{m} \dot x^{m} - {1\over 2} P_{m} P^{m} 
- w_{L\a} \dot \l^{\a}_{L} - \dot \l^{\a}_{R} w_{R \a} \Big)\,,
\ee 
which is BRST invariant and invariant under the gauge transformation (\ref{conD}) if the spinors 
$\l^{\a}_{L}, \l^{\a}_{R}$ are pure. The action is also invariant under supersymmetry 
transformations generated by 
$Q_{\e} = \e^{\a}_{L} q_{L\a} + \e^{\a}_{R} \, q_{R \a}$
where
\be\label{conF}
q_{L\a} = p_{L\a} - {i\over 2} P_{m} (\g^{m} \t_{L})_{\a}\,, ~~~~~
q_{R\a} = p_{R\a} - {i\over 2} P_{m} (\g^{m} \t_{R})_{\a}\,,
\ee
which anticommute with the BRST operators $Q_{L}$ and $Q_{R}$. 

The physical states are identified with the BRST cohomology 
at ghost number 1 and the cohomology is computed 
by the following equations
\be\label{cohoA}
Q_{L} | \psi > =0 \,, Q_{R} |\psi> =0\,, ~~~~~~
|\psi> \neq Q_{L} |\Omega_{L}> +  Q_{R} |\Omega_{R}>
\ee
with $Q_{R} |\Omega_{L} > = Q_{L} |\Omega_{R}> =0$. 
The physical state $|\psi>$ has ghost number one and 
the parameters of the gauge transformations $|\Omega_{L/R}>$ have ghost number 
zero. The states $|\psi>$ are 
obtained by acting with normal-ordered combinations 
of operators $x^{m}, \theta^{\a}, \dots....$ on the vacuum $|0>$. 
The complete analysis of eqs.~(\ref{cohoA}) in d=10 N=2 case 
has been given in \cite{Berkovits:2001ue,Grassi:2004ih}, and based 
on those results it can be  shown that the solution of these equations yields 
the off-shell multiplet of  N=2 d=4 supergravity.  


\subsection{Coupling the superparticle to the background}
\label{background}

As illustrated in \cite{Berkovits:2002uc} the superparticle 
$N=2$ can be coupled to a $N=2$ supergravity background. 
The deformation of the action $S + \int d\tau V$ has to be BRST invariant in order to define 
gauge invariant correlation functions. For constant backgrounds, 
the BRST invariant action is given by 
\be\label{conFAA}
S_{R} = S_{0} + \int d\tau \,\Big( P^{m} g_{mn} P^{n} + 
B_{mn} L^{mn} + 
\Psi^{\a}_{Lm} q_{L\a} P^{m} + 
+ q_{R\bar\a} P^{m} \Psi^{\a}_{Lm} +
i\, q_{L\a} F^{\a \b} \, q_{R \b}\Big) \,,
\ee
where $g_{mn}$ is the usual metric deformation, $B_{mn}$ is the NS-NS 
two form, $\Psi^{\a}_{Lm}$ and $\Psi^{\a}_{Rm}$ are the gravitinos and 
$F^{\a \b}$ are the R-R field strengths. 
The fields $P_{m}, q_{L\a}, q_{R\bar\a}$ and 
\be\label{conFB}
L^{mn} = P^{[m} x^{n]} + 
{1\over2} p_{L} \g^{mn}\t_{L} + {1\over2} p_{R} \g^{mn}\t_{R} + 
{1\over2} w_{L} \g^{mn} \l_{L} + {1\over2} w_{R} \g^{mn}\t_{R} \,,
\ee
are BRST invariant. As a consequence, the action (\ref{conFAA}) is 
invariant if the backgrounds are constant. 
Given that, we can obtain the action (\ref{conFAA}) 
by an equation similar to (\ref{conDA}) 
(see \cite{Oda:2001zm,Berkovits:2002uc}). 
In the following we will set all background fields to zero except for the RR 
graviphoton and the metric $G_{mn}$ and we will take them to be constant.  	

The advantage of working in 4 dimensions is due to weaker constraints to 
which the background has to satisfy and the absence of backreaction. 
Indeed, as is been shown in \cite{Grassi:2005sb}, in d=4 the 
BRST cohomology implies only that the background fields belong to 
off-shell supermultiplets and no equations of motion are necessary 
(see also \cite{Morera:2005jz}). In the case of closed superstrings 
\cite{Grassi:2005gt}, it can be shown that the BRST conditions implies 
only some kinematical restrictions on the background. For that 
reason, one can choose suitable background enforcing the absence of the 
backreaction.


\section{RR fields and Non-commutative Superspace I}
\label{RRI}

Setting the background fields $B_{mn}$ and the gravitinos $\Psi^{\a}_{Lm}$, $\Psi^{\bar\a}_{Rm}$ 
to zero and we assume that $F^{\a \b} = F^{\b \a}$, we obtain the new action
$$
S_{R}= \int d\tau 
\Big[ \dot \t^{\a}_{L} p_{L\a} + \dot \t^{\a}_{R} p_{R\a} + 
i\, \Big(p_{L\a} - {i\over 2} P_{m} (\g^{m} \t_{L})_{\a}\Big) F^{\a \b} 
\Big(p_{R \b} - {i\over 2} P_{m} (\g^{m} \t_{R})_{\b}\Big) \Big]
$$
\be\label{conG}
+ \int d\tau \Big[
P_{m} \dot x^{m} - {1\over 2} G_{mn} P^{m} P^{n} - 
w_{L\a} \dot \l^{\a}_{L} - \dot \l^{\a}_{R} w_{R \a} \Big]\,,
\ee
where $G_{mn} = \eta_{mn} + g_{mn}$. The presence of RR fields breaks the supersymmetry. The amount 
of supersymmetry preserved in this background is given by the equations
\be\label{conH}
P_{m} F^{\a\b} \g^{m}_{\b \g} \e_{R}^{\g} =0\,, ~~~~~ 
P_{m} F^{\a \b} \g^{m}_{\a \g} \e_{L}^{\g} =0\,.
\ee
which are the usual Killing equations for spinors if one redefines the 
supersymmetry parameters with $\e'_{L} = \not\!P \e_{L}$ and $\e'_{R} = \not\!P \e_{R}$, for 
off-shell momentum $P_{m}$. 
No contribution is added to the ghost action and this simplifies the analysis. 

From  the action (\ref{conG}) we can derive the equations of motion for 
$\t^{\a}_{L}$ and $\t_{R}^{\a}$
\be\label{conI}
- \dot \t^{\a}_{L} +i\, F^{\a \b} p_{R \bar\b} + {1 \over 2} (F\!\not\!P)^{\a}_{~\b} \t_{R}^{\b} =0 \,, ~~~~~~
- \dot \t^{\a}_{R} - i F^{\b \a} p_{L \b} - {1 \over 2} (F\!\not\!P)_{\b}^{~\a} \t^{\b}_{L} =0 \,,
\ee
which can be solved in terms of $p_{L\a}$ and $p_{R\bar\a}$. We assume for the time being that 
$F^{\a \b}$ is an invertible matrix. On the contrary, if $F$ is not invertible on the spinor space, one 
can decompose any spinor into a part belonging to $\ker(F)$ and to $\ker(F)^{\perp}$. The spinors 
belonging to the kernel of F do not enter the coupling term in~(\ref{conG}) 
and therefore can be treated separately, in that case there is a residual supersymmetry. 

Notice that the RR field $F^{\a\b}$ plays the role of the mass in the 
case of non-relativistic charge particle in Sec.~\ref{particle}. Therefore, 
we are interested in studying the limit $||F|| \rightarrow \infty$, where 
$|| \cdot ||$ means a measure of the intensity of the RR field 
strength.\footnote{The relation between lowest Landau levels and RR 
fields is explored in  \cite{Hatsuda:2003ry}} Moreover, the RR 
fields does not seem to play the role of the magnetic field $B_{mn}$ of 
Sec.~\ref{particle}. Indeed, the anticommutation relations among the fermionic 
constraints $d_{\a L}$ and $d_{\a R}$ do not contain the field $F^{\a\b}$ 
in contrast to the corresponding constraints $\Pi^{m}$ of the bosonic case. 

To exploit the analogy between the mass term of (\ref{ppA}) and 
the RR-dependent terms of (\ref{conG}), we further manipulate  
the action.  
Substituting (\ref{conI}) in the action (\ref{conG}), one obtains
\be\label{conL}
S_{R} = \int d\tau 
\Big[ i \dot \t^{\a}_{L} F^{-1} _{\a \b} \dot \t^{\b}_{R} + {i \over 2} \t_{L} \not\!P \dot\t_{L} + 
{i\over 2} \dot\t_{R} \not\!P \t_{R} 
\Big] + 
\int d\tau \Big[ P^{m} \dot x^{m} - {1\over 2} P^{m} P_{m} -
w_{L\a} \dot \l^{\a}_{L} - \dot \l^{\a}_{R} w_{R \a} \Big]\,. 
\ee
The first term is a kinetic term for the fermions which is 
quadratic in the derivatives. This is an usual term for spinors, but it is 
always present in string models in superspace. The second and the third term in the 
action resemble the spinorial part of the Brink-Schwarz action. 

This action suggests that this is a superparticle moving on a supergroup manifold 
with coordinates $x^{m}, \t_{L}^{\a}$ and $\t^{\bar \a}_{R}$. Denoting by 
${\cal Q}_{L \a}, {\cal Q}_{R \bar\a}$ and 
by 
${\cal P}_{m}$ the abstract generators of the algebra 
$\{{\cal Q}_{L\a}, {\cal Q}_{L\b}\} = -\g^{m}_{\a\b} {\cal P}_{m}$ and $\{{\cal Q}_{R\a}, {\cal Q}_{R\b}\} = -\g^{m}_{\a\b} 
{\cal P}_{m}$,  
we find the following MC forms
\be\label{MC}
g^{-1} d g = 
(d x^{m} + d\t_{L} \g^{m} \t_{L} + d \t_{R} \g^{m} \t_{R}) 
{\cal P}_{m} + 
d \t^{\a}_{L} {\cal Q}_{L\a} + d \t^{\a}_{R} {\cal Q}_{R \a}\,,
\ee 
and the metric for the algebra have the following non-vanishing entries
\be\label{alA}
({\cal P}_{m}, {\cal P}_{n} ) = g_{mn}\,, ~~~~~
({\cal Q}_{R \a}, {\cal Q}_{L \b} ) = F^{-1}_{\a \b}\,.
\ee
Notice that the MC forms are not supersymmetric invariant since the supersymmetry is 
broken by the presence of the RR fields. Eliminating $P_{m}$ from the action (\ref{conL}) and 
using the metric given in (\ref{alA}), the action (\ref{conL}) 
can be written as 
\be\label{newA}
S_{R} = \int d\tau \Big[ (g^{-1} \dot g, g^{-1} \dot g) 
- w_{L\a} \dot \l^{\a}_{L} - \dot \l^{\bar\a}_{R} w_{R \bar\a} \Big]\,. 
 \ee
This is similar to the result found in \cite{Berkovits:1999im}, where the authors showed that in the case 
of $AdS_{3}\times S^{3}$, one finds a sigma model on a supergroup. It is interesting that the same 
situation is reproduced in the present context (it should also be possible to do it for string theory in 
10 dimension with constant RR fluxes as studied in 
\cite{Berkovits:2000yr,Berkovits:2002zv,deBoer:2003dn}, but 
nobody found a convenient set of variables yet).

After the conjugated momenta $p_{\a L/R}$ are removed, the equations of motion for the spinors read 
\be\label{spinA}
\ddot{\t}^{\a}_{L} + {1\over 2} F^{\a\bar\b} \not\!P_{\bar\b \bar\g} \dot\t^{\bar\g}_{R} =0\,, ~~~~~~
\ddot{\t}^{\bar\a}_{R} - {1\over 2} \dot\t^{\g}_{L}  \not\!P_{\g \b} F^{\b \bar\a} =0\,. ~~~~
\ee
Since on-shell we have $\dot P_{m} =0$ and since we choose constant RR field strengths, 
we can integrate once the above equations 
to get
\be\label{spinB}
\dot{\t}^{\a}_{L} + {1\over 2} F^{\a\bar\b} \not\!P_{\bar\b \bar\g} \t^{\bar\g}_{R} =C^{\a}_{L}\,, ~~~~~
\dot{\t}^{\bar\a}_{R} - {1\over 2} F^{\bar\a \b} \not\!P_{\b \g} \t^{\g}_{L}  =C^{\bar\a}_{R}\,. ~~~~
\ee
where $C_{L/R}$ are integration constants to be fixed by boundary conditions 
(in order to avoid any new constant non-covariant quantity we choose to set them to zero). 
Inserting (\ref{spinB}) into (\ref{spinA}), 
we arrive at the decoupled equations for $\t^{\a}_{L}$ and $\t^{\bar\a}_{R}$
\be\label{spinC}
\ddot{\t}^{\a}_{L} - {1\over 4} (F \not\!P F^{T} \not\!P)^{\a}_{~\b} \t^{\b}_{L} = 0 \,,~~~~~
\ddot{\t}^{\bar\a}_{R} - {1\over 4} \t^{\bar\b}_{R} (\not\!P F^{T} \not\!P F)_{\bar\b}^{\bar\a} = 0 \,.
\ee
These equations show that the RR fields play the role of a mass term for the fundamental fields 
$\theta^{\a}$ and the matrix $(F\!\not\!\!PF^{T}\!\!\not\!\!P)$ is constant. This is a well-known phenomena in 
$pp$-waves background \cite{Metsaev:2001bj,Metsaev:2002re}: 
the spectrum of worldsheet theory becomes massive. In addition, we can  
see that the theory is not supersymmetric: the bosonic partner is not massive.\footnote{It has to be recalled that the mass parameter in a curved space does not carefully measure the masslessness of the field. The best 
way to reveal a supersymmetry breaking is to analyze the Killing spinor 
equations in a curved background.}

 Nevertheless, if the matrix \-
$(F\!\not\!\!P F^{T}\!\not\!\!P)$ has some zero eigenvalues, along those directions we recover a partial supersymmetry. 

Let us study two interesting limits: $||F||\rightarrow 0$ and $||F|| \rightarrow \infty$. The first 
limit is the regime where perturbation theory can be used to perform worldline computations. The 
second limit is certainly more interesting due to the fact that very little is know about string theory 
in the presence of strong background fields.  

In the limit $|| F || \rightarrow \infty$, where the norm $|| \cdot ||$ is properly defined, 
the first term can be neglected, and the action is invariant under 
a new $\kappa$-symmetry (notice that 
we have replaced the classical $\kappa$-symmetry of the action (\ref{CBS}) 
with the BRST symmetry given in eq.~(\ref{conC})) 
\be\label{kapA}
\delta_{k} x^{m} = \t_{L} \g^{m} \not\!P \kappa_{L} + \t_{R} \g^{m} \not\!P \kappa_{R} + \zeta P^{m}\,, ~~~~
\ee
$$
\delta_{k} \t^{\a}_{L} = - (\not\!P \kappa)^{\a}_{L} + \zeta \dot\t^{\a}_{L}\,, ~~~~~
\delta_{k} \t^{\bar\a}_{R} = - (\not\!P \kappa)^{\bar \a}_{R} + \zeta \dot\t^{\bar\a}_{R}\,,
$$
where the diffeomorphism ghost $\zeta$ compensates the gauge choice 
$e=1$ by choosing the following solution $\zeta = - \int^{\tau} d\tau' (\dot\t^{\a}_{L} \kappa_{L\a} + \dot\t^{\bar\a}_{R} \kappa_{R\bar\a})$. 

Notice also that from eqs.~(\ref{conI}) it turns out that to take properly the limit of $||F|| \rightarrow \infty$ 
one has to impose the constraints 
\be\label{kapB}
q_{\a L} \approx 0\,, ~~~~~~~ q_{\a R} \approx 0\,. 
\ee
These constraints generate the $\kappa$-symmetry (\ref{kapA}) with the opposite 
sign is front of the transformation rules for the spinors. In addition, they 
include also second-class constraints.  
The theory is quantized and therefore we can use the technique 
discussed in \cite{Brink:1980nz,Brink:1981nb,Schwarz:1982pf}. This yields  
the same non-(anti) commutative superspace which is a consequence of the structure of 
second-class constraints of $q_{\a L/R}$ and it does not depend upon the RR field. Applying the 
Dirac procedure, one finds that 
\cite{Brink:1981nb,Schwarz:1982pf}  
\be\label{DpA}
\Big\{ \t_{R}^{\a} ,  \t^{\b}_{R} \Big\}_{D} \sim \gamma_{m}^{\a\b} x^{m}\,, ~~~~~
\Big\{\t_{L}^{\a} , \t^{\b}_{L}  \Big\}_{D}  \sim \gamma_{m}^{\a\b} x^{m}\,, ~~~ 
\Big\{\t_{L}^{\a} , \t^{\b}_{R}  \Big\}_{D} =0\,.
\ee
This result for the superparticle 
is quite different from the result of superstrings \cite{deBoer:2003dn,Ooguri:2003qp,Seiberg:2003yz} given 
in (\ref{ac}); 
for the superparticle, the anticommutation relations of the Grassmann coordinates (\ref{DpA}) 
are related to the bosonic coordinates and not the RR field. This is due to the fact   
that we have only one set of free parameters, namely $F^{\a\b}$, which have to be 
interpreted as a mass matrix and not as a ``magnetic field''. 

Moreover, for highly curved space one 
has to take into account the radiative corrections to the action (\ref{conL}) 
before taking the limit $|| F || \rightarrow \infty$. 
In fact, as we shall show it below, at one loop there are new pieces generated 
by radiative correction at one-loop 
in the worldline. 

On the other hand, in the limit $|| F || \rightarrow 0$, the first term becomes dominant over the other fermionic terms 
(for the bosonic terms in (\ref{conL}), 
one can also add a background metric $- {1\over 2} g^{mn} P_{m} P_{n}$ and therefore they
cannot be neglected) and the action (\ref{conL}) 
with background metric $g_{mn}$ reduces to 
\be\label{conLA}
S_{R} = \int d\tau 
\Big[ - \dot \t^{\bar \a}_{R} F^{-1} _{\bar \a \b} \dot \t^{\b}_{L}  + 
{1\over 2} g_{mn} \dot x^{n} \dot x^{m}  + 
w_{L\a} \dot \l^{\a}_{L} + \dot \l^{\bar\a}_{R} w_{R \bar\a} \Big]\,. 
\ee
We conclude that we can use the RR background to set up a perturbation theory around weak RR 
backgrounds. Namely, we can consider the action (\ref{conLA}) 
as the quadratic part of the action from which the propagators can be computed, 
and the rest has to be considered 
as a perturbation. 


\subsection{Radiative Corrections}
\label{RCor}

It is easy to compute the radiative corrections to the bosonic inverse 
propagator $\langle x^{m}(\tau) x^{n}(0) \rangle$. 
By computing the free propagators of $x^{m}$ and the off-diagonal propagator of $\t_{L}$ and $\t_{R}$, 
one can obtain the one-loop contribution
\be\label{loopA}
G^{mn}(\tau) =
-{1\over 16} \g^{m}_{\a\b} F^{\b\bar\g} \g^{n}_{\bar\g \bar\d} F^{\a\bar\d} 
\int_{0}^{1} dt |\tau -t| \, |t| 
= - {1\over 16}{\rm Tr}( \g^{m} F \g^{n} F) {\cal P}(\tau)\,,
\ee
where ${\cal P}(\tau)$ is a polynomial of third order in $\tau$. In the 
same way the one-loop corrections to the off-diagonal inverse propagator 
 $\langle \t^{\a}_{L} \t^{\bar \b}_{R} \rangle$
\be\label{loopB}
G_{\a\bar \b}(\tau) = 
-{1\over 16} \g^{m}_{\a\d} F^{\d\bar\g} \g^{n}_{\bar\g \bar\b} g_{mn} {\cal P}(\tau)\,.
\ee
Form this computation, we see how the presence of RR backgrounds in the 
$\langle \t^{\a}_{L} \t^{\bar \b}_{R} \rangle$ leads to modifications in all the couplings and therefore 
the analysis at string coupling $|| F || \rightarrow \infty$ cannot be performed without taking into account 
the radiative corrections. 
It may be possible to re-sum all the contributions. Notice for example that 
the spinors $\t_{L}$ and $\t_{R}$ appear only quadratically in (\ref{conLA}), therefore one can integrate over those 
fields obtaining the following determinant
\be\label{loopC}
\det \left(\begin{array}{cc}
{i \over 2} \not\!P_{\a\b}\p_{\tau} &~~~~~~~ {1\over 2} F^{-1}_{\a\bar \b}\p^{2}_{\tau} \cr
 {1\over 2} F^{-1}_{\bar\a  \b}\p^{2}_{\tau} &~~~~~~~ {i \over 2} \not\!P_{\bar\a\bar\b}\p_{\tau} \end{array} \right)\,.
\ee
The next step will be to integrate over the field $P_{m}$. We will not pursue this analysis here and 
we refer to a subsequent publication \cite{Grassi:2005gt}. 


\section{D0-branes and the BRST symmetry}
\label{d0brane}

Since we have seen that the introduction of the RR fields in the case of 
N=2 d=4 (or d=10) superparticle does not really help for deriving the 
anticommutation relation (\ref{ac}) for the Grassmann coordinates, we need 
to use a different type of superparticle to do it. 

As is known, in string theory there are solitonic degrees of freedom which 
are known as Dp-branes \cite{Polchinski:1998rq}. A given Dp-brane has a 
worldvolume which has $p+1$ dimensions. They are characterized by the 
coupling to the RR fields of the superstrings, and they can be 
described by a low energy effective action which is the sum of a Born-Infeld action 
and a Wess-Zumino term. Among the Dp-branes, 
we can consider the D0-brane which is a particle (the worldvolume is 1-dimensional) 
described by the effective action
\be\label{dA}
S = \int d\tau ( P_{m} \Pi^{m} - {e\over 2} (P_{m} P^{m} + f^{2})) + \int d\tau f_{\a\b}  \theta^{\a}_{L} \dot \theta^{\b}_{R}
\ee 
in the first order formalism with
\be\label{dB}
\Pi^{m} = \dot{x} + {i\over 2} \t_{L} \g^{m} \dot \t_{L} + {i \over 2} \t_{R} \g^{m} \dot \t_{R}\,.
\ee 
and $f_{\a\b}$ is constant. (In the case of d=10 action,  $f_{\a\b}$  is a scalar proportional to $\delta_{\a}^{~\b}$.) The first term is the Born-Infeld term written 
in the first order formalism. The second term is a Wess-Zumino term 
and the coefficient $f_{\a\b}$ is related to the brane tension \cite{Aganagic:1996nn}. 
Indeed it can be viewed as the mass of the $D0$-brane

The action is invariant under $\kappa$-symmetry which yields the supersymmetry of the 
$D0$-brane and it can be quantized using the BRST technique discussed above in Sec.~\ref{quantization}. 
An objection to this might be: the $D0$-brane is not a fundamental degree of freedom and 
there is no need of quantizing the it! However it is shown in  \cite{Anguelova:2003sn} that 
the BRST based on Pure Spinor formulation replaces the $\kappa$-symmetry 
and  provides a guideline how 
to coupled the $D0$-brane to a supergravity background as in the case of the N=2 superparticle. 

In analogy with eqs.~(\ref{conF}), we derive the supersymmetry generators 
$q_{\a, L}$ and $q_{\a,R}$ for the $D0$-brane.
\be\label{dC}
q_{L\a} = p_{L\a} - {i\over 2} P_{m} (\g^{m} \t_{L})_{\a} + f_{\a\b} \t^{\b}_{R}\,, 
\ee
$$
q_{R\a} = p_{R\a} - {i\over 2} P_{m} (\g^{m} \t_{R})_{\a} + f_{\a\b} \t^{\a}_{L}\,.
$$
They couple to the RR field of the supergravity background and the supersymmetry 
is broken because of the presence of RR fields, or equivalently to the presence 
of $D0$-branes.

\section{RR fields and Non-commutative Superspace II}
\label{RRII}

Finally, we can couple the $D0$-brane to RR field of the supergravity background and 
this introduces a new term of the form 
\be\label{RRIIA}
\int d\tau F^{\a\b} q_{\a L} q_{\b R}
\ee
such as in the case of a superparticle coupled to RR fields. Repeating 
the derivation as in sec. \ref{RRI}, we find that 
in the limit of $|| F || \rightarrow \infty$, we need the constraints 
\be\label{RRIIB}
q_{\a L} \approx 0\,, ~~~~~~~ q_{\a R} \approx 0\,. 
\ee
where $q_{\a L}$ and $q_{\a R}$ are given by (\ref{dC}) and they depend on the 
RR field $f_{\a\b}$ generated by the $D0$-brane. Applying the Dirac 
procedure and using the canonical brackets 
$[p_{\a L/R}, \theta^{\b}_{L/R}] = i \delta_{\a}^{~\b}$, we end up with 
the commutation relations
\be\label{RRIIC}
\{\theta^{\a}_{L}, \theta^{\b}_{R}\}_{D} = (f^{-1})^{\a\b}\,,
\ee 
which finally gives the non-(anti)commutation relations between the fermionic coordinates. 

\vspace{.5 cm}

To conclude, we have shown that the coupling of the RR fields and the Wess-Zumino term, in the case of superparticle and $D0$-brane, are fundamental to generate a deformation of anti-commutation 
relations among Grassmann coordinates of the superspace. 
It is point out that the RR fields of the supergravity 
background cannot generate the wanted commutation relations, but it replaces the role of the mass as in the case of the non-relativistic particle moving in a magnetic field. Furthermore, it is shown that the brane tension together with a 
supergravity background yields the wanted commutation relations (\ref{ac}) obtained also in string theory. The role of the mass and and of the RR fields is inverted: in the case of massive bosonic charge particle, we take the limit $m \rightarrow 0$ and 
we derive (\ref{IA}) deformed by $B_{mn}$, in the case of superparticles, we take the limit $|| F|| \rightarrow \infty$ and we derive (\ref{ac}) deformed by the brane tension $f$. We argued that this limit can be taken only by neglecting 
the radiative corrections and a deeper analysis will be presented elsewhere 
\cite{Grassi:2005gt}. 

\section*{Acknowledgements} 

The author would like to thank the organizers and the 
directors of the 43$^{rd}$ International School on Subnuclear Physics held 
in Erice, Sicily in Sept. 2005 where a lecture based on the present paper has been delivered. 
We also thank I. Bars, L. Castellani, A. Lerda, F. Morales, 
Y. Oz, L. Tamassia and N. Wyllard for useful discussions on related subjects and comments. We thank P. Aschieri for a useful discussion based 
on work \cite{Aschieri:2005zs} where a wide class of superspace non-commutative deformations are discussed and that stimulated the present derivation of the anti-commutation relations. A special thank to L. Anguelova for a careful reading of the 
manuscript and for useful suggestions.


\end{document}